\documentclass[12pt,prd]{revtex4}
\usepackage{epsfig}

\newcommand{\be}{\begin{equation}}
\newcommand{\ee}{\end{equation}}
\newcommand{\bea}{\begin{eqnarray}}
\newcommand{\eea}{\end{eqnarray}}
\newcommand{\gsim}{\lower.7ex\hbox{$\;\stackrel{\textstyle>}{\sim}\;$}}
\newcommand{\lsim}{\lower.7ex\hbox{$\;\stackrel{\textstyle<}{\sim}\;$}}

\newcommand{\q}{\emph{quaternion}}

\begin{document}

\title{
On Symmetry Properties of Quaternionic Analogs of Julia Sets}

\author{
A. A. Bogush, A. Z. Gazizov, Yu.\ A. Kurochkin and V. T. Stosui }
\thanks{gazizov@dragon.bas-net.by}
\affiliation{B. I. Stepanov Institute of Physics of the National
Academy of Sciences of Belarus, F. Skariny Ave.\ 70, 220072
Minsk, Belarus}

\begin{abstract}
By means of theory group analysis, some algebraic and geometrical
properties of quaternion analogs of \emph{Julia} sets are
investigated. We argue that symmetries, intrinsic to quaternions,
give rise to the class of identical \emph{Julia} sets, which does
not exist in complex number case. In the case of quadratic
quaternionic mapping $X_{k+1} = X_k^2 + C$ these symmetries mean,
that the shape of fractal \emph{Julia} set is completely defined
by just two numbers, $C_0$ and $|{\bf C}|$. Moreover, for given
$C_0$ the vector part of the \emph{Julia} set may be obtained by
rotation of a two-dimensional \emph{Julia} subset of arbitrary
plane, comprising ${\bf C}$, around the axis ${\bf n} = {\bf
C}/|{\bf C}|$.
\end{abstract}
\maketitle

%%%%%%%%%%%%%%%%%%%%%%%%%%%%%%%%%%%%%

\section{Introduction}
Beautiful and unexpected fractal properties of \emph{Julia-Fatou}
and \emph{Mandelbrot} sets, spawned by a simple iteration rule
\be%
\label{rule} z_{k+1} = z_k^2 + C,
\ee%
with $z$ and control parameter $C$ being complex numbers, have
been intensively studied by many authors \cite{Ma,Pe,Ya}.
Iterations of arbitrary starting point $z_{0}$ in accordance with
(\ref{rule}) ultimately result in the confinement of all $z_k,
k\ge N_{max} >> 1$ in a finite region, a basin of attraction, of
some $z_i$. These $z_i$ are called attractors; they are
completely defined by control parameter $C$.

A key concept of fractal set research is the special case of
$z=\infty$ attractor. In fact, each $C$ of (\ref{rule})
classifies all points of the complex plane as belonging to either
\emph{runaway} subset or \emph{prisoners} one. If $z_0$ belongs
to \emph{runaway} subset, then (\ref{rule}) leads to $z=\infty$.
In the other case, ultimate cycles of $z_k$ will reside in a
basin of some finite attractor.

A fascinating boundary of the basin of attraction of $z=\infty$
is called \emph{Julia} set. All information about shapes and
topologies of these sets is encoded in $C$. These sets are mostly
fractals.

Prisoners of
\be%
\label{Mandl} C_{k+1} = C_k^2 + C_k
\ee%
(i.e.\ of quadratic mapping (\ref{rule}) with starting point $z_0
= 0$) form the famous \emph{Mandelbrot} set. It classifies all
\emph{Julia} sets as either \emph{connected} or \emph{unconnected}
ones. The latter are known as \emph{Cantor dust} or \emph{Fatou}
sets.

There were many attempts to apply fractal sets to various fields
of physics (see \cite{Sm,Us}). It was clear from the very
beginning, that many attractive features of these sets were due to
remarkable properties of algebra of complex numbers. This suggests
to generalize the quadratic mapping (\ref{rule}) to different
algebras and to study their behavior.

In paper \cite{Ga} such a generalization was undertaken for
\emph{double numbers}, which differ from complex ones just in
definition of the imaginary unit $\varepsilon$, i.e.\
$\varepsilon^2 = 1$. Replacement of complex numbers by double
numbers actually means a transition from the Euclidean plane to
the pseudoeuclidean one, the latter being of special interest for
physics, in particular for relativistic kinematics. Another
example of validity of double numbers give two-dimensional
relativistic models, which are prevalent in modern field
theories. They also may be successfully described in terms of
double numbers \cite{BGK}.

Generally speaking, all \emph{hypercomplex algebras}, i.e.\
algebra of double and dual numbers, of quaternions, biquaternions
and of octanions, have  proved to be very convenient in numerous
physical applications. It may be explained by their close
relations with geometries of Euclidean  and pseudoeuclidean
spaces and with spaces of constant curvature \cite{Be}.

In this paper we focus on quaternion algebra. Really, the
\emph{'quaternion language'} seems to be especially natural for
physics of our four-dimensional space-time. It has been already
successfully exploited in describing of rigid body motion
\cite{Bran}, in searching for instanton solutions of Yang-Mills
equations \cite{FrUl}, in the problem of supersymmetric
oscillator \cite{Bo} and many others.

\section{Quaternions}
Let us first briefly remind the fundamentals of quaternion
algebra. The \q\ $X$ is a set of $4$ real numbers $x_0, x_1, x_2,
x_3$ with $3$ imaginary units $i,j,k$. The definition and
commutation properties are as follows:
\be%
\label{qdifin}%
 X = x_0 + i x_1 + j x_2 + k x_3,
\ee%
where
\bea%
i^2  & =  \;\;\;\: j^2  = & k^2 = -1;  \nonumber  \\
i j  & =          -j i  = & k;                    \\
k i  & =          -i k  = & j;         \nonumber  \\
j k  & =          -k j  = & i;         \nonumber
\eea%
Hence \q s are non-commutative with respect to multiplication, so
that
\be%
X_1 \cdot X_2 \neq X_2 \cdot X_1.
\ee%

In some applications a representation of \q\ (\ref{qdifin}) as a
pair of complex numbers,
\be%
X = (x_0 + ix_1)+ (x_2 + ix_3)j,
\ee%
may turn to be useful. The traditional notation of arbitrary
quaternion (\ref{qdifin}) as a combination of scalar $x_{0}$ and
three-dimensional vector ${\bf x}$ parts,
\be%
\label{quatnot}%
X = x_{0} + {\bf x},
\ee%
may be very convenient in many problems. In terms of
(\ref{quatnot}), the operation of quaternion conjugation is simply
\[%
\bar{X} = x_{0} -{\bf x}.
\]%
The product of two arbitrary elements of the quaternionic algebra
(i.e.\ of two four-dimensional vectors) in the traditional
notations reads:
\be%
\label{qmultipl}%
XX^{\prime} = x_{0}x_{0}^{\prime} - ({\bf x}  \cdot {\bf
x^\prime})+
 x_{0} {\bf x^\prime}+
x_{0}^{\prime} {\bf x}+[{\bf x}{\bf x^\prime}].
\ee%
Consequently, the square of any quaternion $X$ looks like
\be%
\label{qsquare}%
 X^2 = XX = x_0^2 - |{\bf x}|^2 + 2 x_0 {\bf x}.
\ee%

It should be noted, that operation of division by nonzero element
($q \neq 0$) is well defined for \q s.

The set of quaternions forms a group with respect to the
operations of addition and multiplication.

\section{Quaternionic analogs of fractal sets}
The above-mentioned algebraic properties of quaternions permit to
introduce a quaternionic analog of the original complex number
\emph{Julia-Fatou} algorithm (\ref{rule}). Namely,
\begin{equation}
X_{k+1}= X^{2}_{k} + C, \label{qrule}
\end{equation}
with $X$ and control parameter $C$ being quaternions. This
iteration rule maps the four-dimensional Euclidean space on
itself. Direct computer experiments justify, that (\ref{qrule})
classifies again all quaternions with respect to $X = \infty$ as
belonging to either \emph{prisoner} or \emph{runaway} subsets.
The border of these subsets is a quaternionic analog of
\emph{Julia} set; it is fully defined by coefficients of mapping
rule, e.g.\ $C$ in case of (\ref{qrule}). However, in contrast to
complex and double numbers, these fractal sets are
four-dimensional.

The first treatment of this problem was given by \emph{A. Norton}
in Ref.~\cite{No}. The author have discussed some aspects of
quadratic mappings
\be%
X_{k+1}= A X^{2}_{k} + B \label{Nrt}
\ee%
with $A, B$ and $X$ being quaternions. It was noted, that the
general case the problem was very difficult both for discussion
and for representation of results.

To simplify the problem, the consideration of Ref.~\cite{No} was
restricted to a special case of coefficients $A$ and $B$
belonging to the complex number subalgebra of quaternions. The
author argued, that this choice allowed to reduce the dimensions
of the resulting \emph{Julia} sets. However, no good arguments
were proposed to support this statement.

A particular attention of Ref.~\cite{No} had been paid to the
analysis of topological properties of quaternionic \emph{Julia}
sets. It was noted, that, due to noncommutativity of quaternions,
there actually exist three different non-equivalent
generalizations of (\ref{rule}). Besides (\ref{Nrt}), one is also
to discuss
\bea%
\label{algnort}%
X_{k+1} &= & X_k A X_k +   B, \\
X_{k+1} &= & X_k^2 A   +   B. \nonumber
\eea%

The main goal of our research is to provide an instrument for
establishing of common features and differences of fractal sets
realized by algorithms (\ref{qrule})--(\ref{algnort}) from
algebraic, theoretical group point of view.

Let us consider transformations connected with multiplications and
additions of quaternions as transformations of the group of motion
of four-dimensional Euclidean space \cite{Be,Fe}. It is easy to
check that (\ref{qrule})--(\ref{algnort}) are invariant under the
following transformations:
\be%
\label{transform}%
X^{'}_{k} = Q X_{k}\bar{Q}, \quad A^{'} = Q A \bar{Q}, \quad B^{'}
= Q B \bar{Q}, \quad C^{'}= Q C \bar{Q},
\ee%
where quaternion $Q$ satisfies the condition
\be%
\label{unitar}%
Q \bar{Q} = 1,
\ee%
i.e.\ $\bar{Q} = Q^{-1}$.

Note, that due to commutativity of complex and double numbers the
analogous transformations, $q = e^{i\varphi}$, mean just trivial
rotation by angle $\varphi$ of the whole plane.

Transformations (\ref{transform}) are inner automorphisms of the
\emph{division ring} of quaternions. They are isomorphic to the
transformations of the group of  three-dimensional rotations
$SO(3.R)$. Note, that $x_0$ and $|{\bf x}|$ stay invariant under
such transformations.

Thus, we actually deal with a class of equivalent iteration rules
\be%
\label{transrule}%
X^{\prime}_{k+1}= A^\prime (X^{\prime}_{k})^2 + B^\prime,
\ee%
which are bound with (\ref{Nrt}) by relations
(\ref{transform}--\ref{unitar}). Applying these transformations to
the mapping (\ref{qrule}), we obtain equivalent rules
\be%
\label{qruleprime}%
X^{\prime}_{k+1}= (X^{\prime}_{k})^2 + C^\prime.
\ee%

To illustrate this symmetry manifestation, let us show, that there
is a freedom in orientation of vector ${\bf C'}$ (remember, that $
C_0^\prime = C_0, |{\bf C'}| = |{\bf C}|$). These transformations
are analogous to the plane transformations of the Lorentz group,
found in Ref.~\cite{Bogush}. For vector parts of quaternions $C$
and $C^\prime$ the plane transformation quaternion $Q$ and its
conjugate $\bar{Q}$ are:
\be%
Q=\frac{\bf{C}+\bf{C^\prime}}{\sqrt{-(\bf{C}+
\bf{C^\prime})^{2}}}\frac{\bar{\bf{C}}}{\sqrt{-\bf{C}^{2}}},
\ee%
\be%
\bar{Q}=\frac{{\bf C}}{\sqrt{-\bf{C}^2}}
\frac{\bar{\bf{C}}+\bar{\bf{C^\prime}}}{\sqrt{-(\bf{C}+
\bf{C^\prime})^{2}}} =\frac{\bf{C}+\bf{C^\prime}}{\sqrt{-(\bf{C}+
\bf{C^\prime})^{2}}}\frac{\bar{\bf{C^\prime}}}{\sqrt{-\bf{C^\prime}^{2}}}.
\ee%

It may be convenient to rotate vector ${\bf C }$ so that to orient
vector ${\bf C'}$ along, say, vector ${\bf i}$:
\be%
C^\prime = C_0 + |{\bf C}|i.
\ee%
If all starting points $X_0$ of (\ref{Nrt}) lie in this complex
plane, the following points $X_k$ will lie in this plane too.
Thus, one obtains a complex number \emph{Julia} set, which is a
subset of the full quaternionic one.

The most important example of the profit of (\ref{transform}) in
studies of quaternionic \emph{Julia} set symmetries is given by
transformations, that leave control parameter $C$ intact, i.e.\
\be%
C^\prime = QC\bar{Q} = C, \;\;\; Q \bar{Q} =1.
\ee%
It is straightforward to check that the proper quaternion is (see
Ref.~\cite{Be}):
\be%
\label{O2}
Q =%
\frac%
{1 + {\bf n} \tan \frac{\varphi}{2}}%
{\sqrt {1 + \tan^2 \frac{\varphi}{2}}}%
= \cos \frac{\varphi}{2} + {\bf n} \sin \frac{\varphi}{2},
\ee%
where ${\bf n} = {\bf C}/|{\bf C}|$.

These transformations form the $O(2)$ group. They are similar to
the gauge symmetries of field theories. One can consider the
resulting \emph{Julia} set as a projective space and bundle
manifold. An invariance of fractal sets under transformation
(\ref{O2}) means that projection of this sets in the three
dimensional subspace are axial symmetric.

These symmetries does not change the zero quaternion component.
They do exist for each $C_O$, but corresponding two-dimensional
Julia subsets are different even for the same $|{\bf C}|$. It
fact, quaternionic symmetries mean, that one needs just two
numbers, $C_O$ and $|{\bf C}|$, to describe all possible shapes of
quaternionic \emph{Julia} sets.

Besides continuous symmetries, \emph{quadratic} mappings of all
hypercomplex algebras also possess a special obvious invariance
under discrete reflections,
\be%
\label{discrete}%
X \leftrightarrow -X.
\ee%
This invariance is due to two-valued property of $X^2$.  Really,
if some point $X_k$ belongs to a \emph{Julia} set, then the same
is true for $-X_k$. This symmetry allows to save the computer
time at calculations of fractal pictures.

It may be interesting to mention, that the special choice of
parameters $A$ and $B$ in Ref.~\cite{No} as complex numbers was
not obligatory. Symmetries (\ref{transform}) imply, that it would
suffice just to orient vector parts of both quaternions in the
same direction. The resulting \emph{Julia} sets would be obtained
by rotation of arbitrary plane \emph{Julia} subset around the
axis oriented along the  direction of their vector parts. The
transverse cross-sections of these \emph{Julia} sets consist of
concentric circles.

\section{Conclusions}
Quaternion algebra provides non-trivial generalizations of usual
\emph{Julia} sets. Although these sets are much more complicated,
the intrinsic quaternionic symmetries allow to simplify the
problem. It turns out that in case of quadratic mapping
(\ref{qrule}) all essentially different quaternionic analogs of
\emph{Julia} sets may be enumerated by just two numbers, $C_0$ and
$|{\bf C}|$. Due to $O(2)$ symmetry, the three-dimensional part
of quaternionic \emph{Julia} sets may be restored by rotation of
some arbitrary two-dimensional (e.g.\ complex number) \emph{Julia}
subset around the axis ${\bf n} = {\bf C}/|{\bf C}|$, that lies
in this plane.

This work was supported by the Belarusian Republic Foundation of
Fundamental Research, project No.\ P-229/585.

%\newpage


\begin{thebibliography}{99}

\bibitem{Ma} Mandelbrot B. B., \emph{The Fractal Geometry of
Nature.} Freeman and company, N.Y., (1982) 468 p.

\bibitem{Pe}  Peitgen H.-O. and Richter P. H., \emph{The Beauty of
Fractals.} Springer-Verlag Berlin-Heidelberg New York Tokio
(1986).

\bibitem{Ya} Yakobson M. V., \emph{The ergodic theory of the one dimensional
maps.} Proc.\ of Dinamical systems-2. VINITI. Ser.\ Modern
problems of mathematics. Fundamental directions. Moscow, Nauka,
{\bf 2} (1985) 204-233. (In Russian).

\bibitem{Sm} Smirnov B. M., \emph{The Problem of the Ball
Lightning.} Moscow, Nauka (1988) 208 p. (in Russian).

\bibitem{Us} Ushenko A. G., \emph{Laser diagnostics of
Biofractals.} \emph{Quantum electronics}, {\bf 29} (1999) 239-245.

\bibitem{Ga} Gazizov A. Z. and Kurochkin Yu.\ A., \emph{The Analog of Julia Sets
Associated with Double Numbers.} Proc.\ of the Int.\ Seminar
"Nonlinear phenomena in Complex Systems", Polatsk (1992) 126-134.

\bibitem{BGK} Bogush A. A., Gritsev V. V. and Kurochkin Yu.\ A., \emph{Analytic
functions of the double variables and theory of relativistic
strings.} Proc.\ of the Int.\ Seminar "Nonlinear phenomena in
Complex Systems", Minsk (1997) 571-573.

\bibitem{Be}
Berezin A. V., Kurochkin Yu.\ A. and Tolkachev E. A.,
\emph{Quaternions in Relativistic Physics.} Minsk, Nauka i
Tekhnika (1989) 198. (In Russian).

\bibitem{Bran}
Branets V. N. and Shmyglevsky I. P., \emph{Application of
Quaternions to the Problems of Rigid Body Orientation.} Moscow,
Nauka (1973) 319 p. (In Russian).

\bibitem{FrUl}
Frid D. S. and Ulenbeck K. K., \emph{Instantons and
Four-Manifolds.} Springer-Verlag Berlin Heidelberg New York Tokyo
(1984) 274 p.

\bibitem{Bo}
Bogush A. A. and Kurochkin Yu. A., \emph{Cayley-Dickson Procedure,
Relativistic Wave Equations and Supersymmetric Oscillators.} Acta
Applicandae Mathematicae, {\bf 50} (1998) 121-129.

\bibitem{No}
Norton A., \emph{Julia Sets in the Quaternions.} Computers and
Graphics, {\bf 13} No.~2 (1989) 266-278.

\bibitem{Fe}
Fedorov F. I., \emph{The Lorentz Group.} Moscow, Nauka (1979) 383
p.

\bibitem{Bogush}
Bogush A. A. and Fedorov F. I., \emph{On transformations of
four-dimensional vectors.} Doclady of the BSSR Academy of
Sciences, {\bf 6} No.~11 (1962) 690-693. (In Russian); \\
Bogush A. A. and Moroz L. G., \emph{Finite Transformations of the
Lorentz Group and Their Representations.} Preprint of the
Institute of Physics of the Academy of Sciences of BSSR, Minsk
(1970) 50 p. (In Russian).

\end{thebibliography}
\end{document}